\documentclass[letterpaper, 12pt]{article}
\usepackage{jheppub}

\usepackage{graphicx}
\usepackage{bbm}
\usepackage{float}
\usepackage{upgreek}
\usepackage{amsmath}
\usepackage{amssymb}
\usepackage{amsthm}
\usepackage{physics}
\usepackage{afterpage}
\usepackage{rotating}
\usepackage{epstopdf}
\usepackage{mathtools}
\usepackage{multirow}
\usepackage{mathrsfs}
\usepackage[labelformat=simple]{subcaption}
\usepackage{calligra}
\usepackage[T1]{fontenc}  
  
\begin{document}

\begin{titlepage}

\setcounter{page}{1} \baselineskip=15.5pt \thispagestyle{empty}
{\flushright {ITP-CAS-24-277}\\}
		
\bigskip\
		
\vspace{0.6cm}
\begin{center}
%{%\fontsize{20}{28}
%{\LARGE \bfseries the\vspace{0.24cm}\\ string}
%{\LARGE \bfseries QCD axion dark matter in the\vspace{0.24cm}\\ dark dimension}
%{\LARGE \bfseries QCD Axion Dark Matter in the\vspace{0.24cm}\\ Dark Dimension}
{\LARGE \bfseries QCD Axion Dark Matter in the Dark\vspace{0.24cm}\\ Dimension}
\end{center}
%\vspace{0.5cm}
\vspace{0.15cm}
			
\begin{center}
{\fontsize{14}{30}\selectfont Hai-Jun Li}
\end{center}
%\vspace{0.1 cm}
\begin{center}
\vspace{0.25 cm}
\textsl{Key Laboratory of Theoretical Physics, Institute of Theoretical Physics, Chinese Academy of Sciences, Beijing 100190, China}\\

\vspace{-0.1 cm}				
\begin{center}
{E-mail: \textcolor{blue}{\tt {lihaijun@itp.ac.cn}}}
\end{center}	
\end{center}
\vspace{0.6cm}
\noindent

The recently proposed dark dimension scenario reveals that axions can be localized on the Standard Model brane, thereby predicting the quantum chromodynamics (QCD) axion decay constant from the Weak Gravity Conjecture: $f_a\lesssim M_5 \sim 10^{9}-10^{10}\, \rm GeV$, where $M_5$ is the five-dimensional Planck mass. 
When combined with observational lower bounds, this implies that $f_a$ falls within a narrow range $f_a\sim 10^{9}-10^{10}\, \rm GeV$, corresponding to the axion mass $m_a\sim 10^{-3}-10^{-2}\, \rm eV$. 
At this scale, the QCD axion constitutes a minor fraction of the total cold dark matter (DM) density $\sim 10^{-3}-10^{-2}$.
In this work, we investigate the issue of QCD axion DM within the context of the dark dimension and demonstrate that the QCD axion in this scenario can account for the entire DM abundance through a simple two-axion mixing mechanism. 
Specifically, we consider the resonant conversion of an axion-like particle (ALP) into the QCD axion. 
We find that, in a scenario where the ALP possesses a mass of approximately $m_A \sim 10^{-5} \, \rm eV$ and a decay constant of $f_A \sim 10^{11} \, \rm GeV$, the QCD axion in the dark dimension can account for the overall DM.
The ALP required within this specific range may originate from the grand unification of gauge forces in the dark dimension.
			
\vspace{2.2cm}
			
\bigskip
\noindent\today
\end{titlepage}
			
\setcounter{tocdepth}{2}
			
\hrule
\tableofcontents			
\bigskip\medskip
\hrule
\bigskip\bigskip
%\pagebreak\
 
%\newpage

\section{Introduction}%%%%%%%%%%%%%%%%%%%%%%%%%%Introduction

The axion is a pseudoscalar field that was initially foreseen by the Peccei-Quinn (PQ) mechanism \cite{Peccei:1977hh, Peccei:1977ur}. 
This mechanism is characterized by a spontaneously broken global $\rm U(1)_{PQ}$ symmetry, wherein the axion, denoted as $\theta$, undergoes a transformation of $\theta \to \theta + c$, with $c$ being a constant. 
The primary objective of this mechanism is to address the strong CP problem in the Standard Model (SM) in a dynamic manner; consequently, it is also referred to as the quantum chromodynamics (QCD) axion \cite{Weinberg:1977ma, Wilczek:1977pj, Kim:1979if, Shifman:1979if, Dine:1981rt, Zhitnitsky:1980tq}.
This axion acquires a small mass from QCD non-perturbative effects \cite{tHooft:1976rip, tHooft:1976snw}. 
When the PQ symmetry is broken by the QCD instanton, it can lead to the induction of an axion potential. 
As a result, the QCD axion can stabilize at the CP conservation minimum, thereby dynamically resolving the strong CP problem \cite{Hook:2018dlk}.

Meanwhile, the QCD axion serves as a natural candidate for cold dark matter (DM), which can be non-thermally produced in the early Universe through the misalignment mechanism \cite{Preskill:1982cy, Abbott:1982af, Dine:1982ah}. 
The oscillations of the axion in its potential can contribute to the overall DM density.
See also refs.~\cite{Marsh:2015xka, DiLuzio:2020wdo, OHare:2024nmr} for recent reviews.
By assuming the $\sim\mathcal{O}(1)$ initial misalignment angle $\theta_i$, the misalignment mechanism imposes an upper limit on the classical QCD axion window, which can be expressed as
\begin{eqnarray}
10^{8}-10^{9}\, {\rm GeV} \lesssim f_a\lesssim 10^{11}-10^{12}\, {\rm GeV}\, ,
\end{eqnarray}
where $f_a$ denotes the QCD axion decay constant. 
In the absence of fine-tuning of the initial misalignment angle, a large $f_a$ would lead to an overproduction of the QCD axion DM, thereby establishing this upper bound.
On the other hand, the lower bound of $f_a$ is approximately $\sim10^{8}-10^{9}\, {\rm GeV}$, which stems from astrophysical observation constraints, including the duration of the neutrino burst from supernova SN\,1987A \cite{Raffelt:1987yt, Turner:1987by, Mayle:1987as} and the cooling rate of neutron stars \cite{Leinson:2014ioa, Hamaguchi:2018oqw, Buschmann:2021juv}, among others.

In general, axions originating from four-dimensional models can be categorically designated as ``type I axions.$"$
Additionally, ``type II axions,$"$ which encompass numerous axion-like particles (ALPs) as well as the QCD axion, can arise from higher-dimensional gauge fields \cite{Witten:1984dg, Green:1984sg, Choi:2003wr}.
In the context of string theory \cite{Svrcek:2006yi, Conlon:2006tq}, lower-dimensional axions can emerge by integrating higher-dimensional gauge fields over cycles within the compactified space. 
The resultant axion decay constant is dictated by the internal geometry of the compactification, often referred to as the model-dependent axion. 
Furthermore, there exists the model-independent axion, which remains steadfastly unaffected by the structure of the internal manifold. 
See also $\rm e.g.$ refs.~\cite{Reece:2024wrn, Choi:2024ome} for recent discussions on axions in extra dimensions.
 
The recently proposed dark dimension scenario \cite{Montero:2022prj} predicts a single large extra dimension with the range $L_5\sim 1-10\, \rm \mu m$, leading to a specific corner of the quantum gravity landscape that corresponds to an asymptotic region of the field space.
This scenario is motivated by the smallness of dark energy, informed by Swampland principles \cite{Vafa:2005ui, Ooguri:2006in}, and constrained by observational data.
In this context, the SM is localized on a codimension-one brane within the five-dimensional spacetime. 
Furthermore, this scenario gives rise to intriguing phenomenology related to the dark dimension, as explored in various studies \cite{Gonzalo:2022jac, Law-Smith:2023czn, Obied:2023clp, Anchordoqui:2022svl, Anchordoqui:2022tgp, Anchordoqui:2022txe, Anchordoqui:2023etp, Anchordoqui:2023tln, Anchordoqui:2024akj, Anchordoqui:2024dxu, Anchordoqui:2024tdj, BitaghsirFadafan:2023hwg, Schwarz:2024tet, Heckman:2024trz}. 
See also ref.~\cite{Vafa:2024fpx} for a recent review.

Axions in the dark dimension have recently been investigated in ref.~\cite{Gendler:2024gdo}. 
Within this context, it is natural to localize the QCD axion on the SM brane. 
By applying the Weak Gravity Conjecture (WGC) \cite{Arkani-Hamed:2006emk} to the QCD axion, we can obtain an upper bound for the axion decay constant, $f_a\lesssim M_5 \sim 10^{9}-10^{10}\, \rm GeV$, where $M_5$ represents the five-dimensional Planck mass.
Additionally, one can derive this inequality by considering axion propagation throughout the entire five-dimensional bulk.
On the observational front, constraints indicate a lower limit for the classical QCD axion window, specifically $f_a\gtrsim 10^{8}-10^{9}\, {\rm GeV}$. 
Consequently, if the dark dimension scenario accurately describes our Universe, the QCD axion localized on the SM brane must possess a narrow range of axion decay constant \cite{Gendler:2024gdo}.  
This range is expressed as follows
\begin{eqnarray}
f_a\sim 10^{9}-10^{10}\, \rm GeV\, ,
\label{DD_f_a}
\end{eqnarray}
which corresponds to the zero-temperature QCD axion mass
\begin{eqnarray}
m_a\sim 10^{-3}-10^{-2}\, \rm eV\, .
\end{eqnarray}
Notably, eq.~\eqref{DD_f_a} falls precisely within the established range of the classical QCD axion window. 
By employing the misalignment mechanism, it can be determined that the QCD axion within this specific range comprises a small fraction of the overall DM density, ranging from $\sim 10^{-3}$ to $10^{-2}$.

In this work, we investigate the issue of QCD axion DM within the context of the dark dimension. 
We demonstrate that the QCD axion in the dark dimension can fully account for the abundance of DM, which is achieved through a straightforward two-axion mixing mechanism. 
The concept of axiverse, a Universe populated by a multitude of axions, including both the QCD axion and a large number of ALPs, has been discussed extensively in the literature \cite{Arvanitaki:2009fg, Acharya:2010zx, Cicoli:2012sz, Demirtas:2018akl, Reig:2021ipa, Demirtas:2021gsq, Alexander:2024nvi}. 
Within this framework, it is plausible to consider the cosmological evolution of multiple axions in the early Universe.
Here, we focus on the resonant conversion of ALP into QCD axion in the dark dimension scenario, a process that can take place prior to the critical temperature of the QCD phase transition.
This similar phenomenon of two-axion mixing has been extensively studied over the past decade, and the resulting QCD axion DM abundance can be either effectively suppressed or enhanced \cite{Hill:1988bu, Cyncynates:2023esj, Li:2024okl, Li:2024kdy, Kitajima:2014xla, Daido:2015cba, Ho:2018qur, Li:2023det, Li:2023xkn, Li:2023uvt, Murai:2024nsp}.
Our findings indicate that, in a scenario where the ALP possesses a mass of approximately $m_A \sim 10^{-5} \, \rm eV$ and a decay constant of $f_A \sim 10^{11} \, \rm GeV$, the resulting QCD axion from the resonant conversion can indeed constitute the entirety of the cold DM abundance.
Lastly, we briefly discuss the adiabatic condition in axion resonant conversion and the possible origins of the ALP with a higher-scale decay constant.

The rest of this paper is structured as follows.  
In section~\ref{sec_review_Axions_in_the_dark_dimension}, we provide a concise overview of the QCD axion within the context of the dark dimension.
In section~\ref{sec_QCD_axion_dark_matter}, we discuss the QCD axion DM in this scenario and demonstrate how the abundance of QCD axion DM is enhanced through the resonant conversion of ALP into QCD axion.
Finally, the conclusion is presented in section~\ref{sec_Conclusion}.

\section{Axions in the dark dimension}%%%%%%%%%%%%%%%%%%%%%%%%%%QCD axion dark matter 
\label{sec_review_Axions_in_the_dark_dimension}

In this section, we first provide a concise overview of axions from higher-dimensional gauge fields.
Subsequently, we briefly review the QCD axion within the context of the dark dimension.

\subsection{Axions from higher-dimensional gauge fields}

Here we introduce axions that originate from higher-dimensional gauge fields.
For a more comprehensive understanding, please refer to recent refs.~\cite{Reece:2024wrn, Choi:2024ome}.

Firstly, we demonstrate that the theory of a higher-dimensional $p$-form gauge field can generate massless four-dimensional axions.
Assume that spacetime consists of $d = (4+n)$ dimensions, which is manifested as a warped product compactification 
\begin{eqnarray}
M = X \times_w Y\, ,
\end{eqnarray} 
where $X$ represents the four-dimensional spacetime, and $Y$ denotes the $n$-dimensional space. 
The metric is given by 
\begin{eqnarray}
ds^2=w(y)ds_X^2+ds_Y^2\, ,
\end{eqnarray} 
where $w(y)\ge0$ is the warping.
If there exists a $p$-form gauge field on $M$, then for every independent non-torsion $p$-cycle in $Y$, there corresponds a distinct massless, periodic four-dimensional axion field.
Considering a $p$-form U(1) gauge field
\begin{eqnarray}
A_p=\dfrac{1}{p!}A_{\mu_1\cdots \mu_p} dx^{\mu_1}\wedge\cdots\wedge dx^{\mu_p}\, ,
\end{eqnarray} 
where $\wedge$ is the wedge product, it includes the local gauge transformations 
\begin{eqnarray}
A_p\to A_p+d\lambda_{p-1}\, , 
\end{eqnarray}
and the large gauge transformations
\begin{eqnarray}
A_p\to A_p+ 2\pi \sum_{i=1}^{b_p(Y)}n_i \omega_p^{(i)}\, , \quad n_i\in \mathbb{Z} \, ,
\end{eqnarray} 
where $b_p(Y)$ represents the Betti number, and $\omega_p^{(i)}$ represents the corresponding cohomology class.
In $d = (4+n)$ dimensions, the standard kinetic term for the $p$-form gauge field can be described by
\begin{eqnarray}
\int_M -\dfrac{1}{2e_p^2}\Phi(x,y) dA_p(x,y) \wedge \star dA_p(x,y)\, ,
\label{S1}
\end{eqnarray}
where $e_p$ represents the $p$-form gauge coupling, $\Phi(x,y)$ denotes the scalar modulus field, and $\star$ is the Hodge star operation, which transforms a $p$-form into a $(d-p)$-form in $d$-dimensional spacetime.
The equation of motion of $A_p$ is given by 
\begin{eqnarray}
d(\Phi \star dA_p)=0\, .
\end{eqnarray}
Assuming a background solution where $\Phi = \Phi(y)$ is independent of $x$, an ansatz for a perturbed four-dimensional axion field around this background is expressed as
\begin{eqnarray}
A_p(x,y)=\theta(x) \omega_p(y)\, ,
\end{eqnarray}
where $\omega_p$ is a $p$-form on $Y$.
In general, considering a collection of such axions, the ansatz generalizes to
\begin{eqnarray}
A_p(x,y)=\sum_{i=1}^{b_p(Y)} \theta_i(x) \widehat{\omega}_p^{(i)}(y)\, ,
\end{eqnarray}
where $\widehat{\omega}_p^{(i)}$ represents the warped harmonic representative. 
Plugging this into eq.~\eqref{S1}, the four-dimensional kinetic term is given by
\begin{eqnarray}
\int_X -\dfrac{1}{2}k_{ij}\, d\theta_i(x)\wedge \star_X d\theta_j(x)\, ,
\end{eqnarray}
with the matrix
\begin{eqnarray}
k_{ij}=\int_Y \dfrac{1}{e_p^2} w(y)\Phi(y)\widehat{\omega}_p^{(i)}\wedge \star_Y \widehat{\omega}_p^{(j)}\, .
\end{eqnarray}
The diagonal entries of the kinetic matrix $k_{ij}$ can be interpreted as the squares of axion decay constants, denoted by 
\begin{eqnarray}
f_i \equiv \sqrt{k_{ij}}\, .
\end{eqnarray}
In the context of large gauge transformations, the redundancy manifests as a precise periodicity constraint on four-dimensional axions, expressed as 
\begin{eqnarray}
\theta_i(x)\to \theta_i(x)+2\pi n_i\, , \quad n_i\in \mathbb{Z} \, .
\end{eqnarray}

Above is the depiction of massless four-dimensional axions originating from higher-dimensional $p$-form gauge fields.
Additionally, there are various effects that contribute to a small mass for these four-dimensional axions, such as their couplings to objects carrying an $A_p$ gauge charge, as well as Chern-Simons (CS) interactions involving $A_p$ with itself or other gauge fields.
The existence of such axions represents a topological characteristic inherent to the theory.
Specifically, there is no requirement for supplementary frameworks such as supersymmetry to elucidate why an axion in an extra dimension is light; it is inherently exponentially suppressed in mass, as long as it does not acquire a tree-level mass contribution from topological terms.
Consequently, these axions alleviate the axion quality problem significantly \cite{Reece:2024wrn}.

\subsection{Axions in the dark dimension} 

In this subsection, we briefly review the QCD axion in the dark dimension.
The dark dimension scenario \cite{Montero:2022prj} predicts one extra mesoscopic dimension in the micron range, which is achieved by applying various Swampland principles to the dark energy.
The dark energy in Planck units $\Lambda$ is a hierarchically small parameter. 
Since the emergence of any small parameters within the framework of quantum gravity necessitates the presence of towers of light states, which can arise due to either some dimensions expanding to a large size or a critical string becoming light \cite{Lee:2019wij}, the sole experimentally allowed model featuring large extra dimensions in this context is one that involves a single large extra dimension with the length
\begin{eqnarray}
L_5\sim \Lambda^{-1/4}\sim 1-10\, \rm \mu m\, ,
\end{eqnarray} 
corresponding to the five-dimensional Planck mass
\begin{eqnarray}
M_5 \sim L_5^{-1/3}M_p^{2/3} \sim 10^{9}-10^{10}\, \rm GeV\, ,
\end{eqnarray} 
where $M_p$ is the Planck mass in our four-dimensional spacetime.
In this scenario, the SM fields should be localized in the fifth dimension, as any deviation would result in the production of a tower of light particles for each SM field --- a phenomenon inconsistent with experimental observations.

The QCD axion within the context of the dark dimension was recently investigated in ref.~\cite{Gendler:2024gdo}. 
That work delves into two scenarios: one where the QCD axion is localized on the SM brane, and another where the axion propagates freely within the bulk of the fifth dimension. 
They find that when the QCD axion is localized on the SM brane, the axion decay constant is subject to an upper bound, as dictated by the WGC:
\begin{eqnarray}
f_a\lesssim M_5 \sim 10^{9}-10^{10}\, \rm GeV\, .
\end{eqnarray}
The WGC \cite{Arkani-Hamed:2006emk} asserts that, for compatibility with quantum gravity, there must exist a U(1) charged particle whose gravitational interaction is weaker than its U(1) gauge interaction. 
In Planck units, this translates to the existence of a ``superextremal$"$ particle that satisfies the following inequality
\begin{eqnarray}
\dfrac{Q}{m} \ge \dfrac{\mathcal{Q}}{M}\bigg|_{\rm ext}=\mathcal{O}(1)\, ,
\end{eqnarray}
where $Q=qg$, with $q$ being the quantized charge and $g$ the gauge coupling, $\mathcal{Q}$ and $M$ represent the charge and mass of an extremal black hole, respectively.
The rationale behind this conjecture is that, in the absence of such a particle adhering to the WGC bound, an extremal charged black hole would be unable to discharge through the Hawking evaporation and would instead decay into a charged remnant, which poses issues with the Bekenstein entropy bound \cite{Bekenstein:1973ur}.
Notice also that in the context of multiple U(1) gauge interactions, there exists a strong version of the WGC \cite{Cheung:2014vva}. For particle species $i$, with charge vector $\vec{q}_i$, mass $m_i$, and defining the vector $\vec{z}_i \equiv \vec{q}_i/m_i$ (in units where an extremal black hole has $|\vec{z}_{\rm BH}|=1$); the mild WGC is equivalent to the fact that the convex hull of the vectors $\vec{z}_i$ contains the unit ball.
For recent reviews on this topic, see $\rm e.g.$ refs.~\cite{Harlow:2022ich, vanBeest:2021lhn, Reece:2023czb, Rudelius:2024mhq}.

The applicability of the WGC extends beyond mere particles, encompassing any charged state associated with a $p$-form gauge field. 
Specifically, for any given $p$-form gauge field, there ought to exist a ($p-1$)-dimensional state, characterized by a charge-to-mass ratio that surpasses or equals that of an extremal ($p-1$) black brane.
When applied to the $0$-form gauge field, namely the axion, we have the inequality
\begin{eqnarray}
S_{\rm inst}\lesssim \dfrac{M_p}{f_a}\, ,
\end{eqnarray}
where $S_{\rm inst}$ represents the instanton action, and $f_a$ is the axion decay constant.  
It is worth noting that, due to the absence of an extremal solution for the instanton, this inequality encompasses an undefined factor of order one. 
Given that the dilute instanton gas approximation necessitates $S_{\rm inst} \gtrsim 1$ to maintain control over the instanton expansion, we can obtain the axion WGC bound
\begin{eqnarray}
f_a\lesssim M_p\, .
\end{eqnarray}
Now, let us consider a four-dimensional brane localized in the fifth dimension and apply the WGC to particles localized on this brane.
This yields 
\begin{eqnarray}
\dfrac{m}{M_p}\le g\, ,
\end{eqnarray}
which can be rewritten by  
\begin{eqnarray}
\dfrac{m}{M_5}\le g M_5^{1/2} L_5^{1/2}\, .
\end{eqnarray}
In the most extreme scenario, where $L_5\sim M_5^{-1}$, the inequality simplifies to
\begin{eqnarray}
\dfrac{m}{M_5}\le g\, .
\end{eqnarray}
In this case, we obtain the axion WGC bound \cite{Gendler:2024gdo}
\begin{eqnarray}
f_a\lesssim M_5\, .
\end{eqnarray}
Alternatively, this inequality can be derived by considering axion propagation throughout the entire five-dimensional bulk. 
Combining this with the observational lower bound on the classical QCD axion window, $f_a\gtrsim 10^{8}-10^{9}\, {\rm GeV}$, the QCD axion localized on the SM brane falls within a narrow range for the axion decay constant
\begin{eqnarray}
f_a\sim 10^{9}-10^{10}\, \rm GeV\, .
\end{eqnarray}
In this context, it is reasonable to consider the QCD axion as the DM candidate, which is the focus of this work.
However, as discussed in \cite{Gendler:2024gdo}, the QCD axion within this range can only contribute a small fraction to the overall DM density. 

\section{QCD axion dark matter in the dark dimension}%%%%%%%%%%%%%%%%%%%%%%%%%%QCD axion dark matter in the dark dimension
\label{sec_QCD_axion_dark_matter}

In this section, we investigate the QCD axion as DM in the dark dimension.
We first review the QCD axion DM through the misalignment mechanism, then we discuss the enhancement of the QCD axion DM abundance within the context of a simple two-axion mixing scenario.
Our aim is to demonstrate that the QCD axion in the dark dimension can fully account for the DM abundance.

\subsection{QCD axion dark matter} 
 
Here we discuss the QCD axion DM in the dark dimension scenario through the misalignment mechanism \cite{Preskill:1982cy, Abbott:1982af, Dine:1982ah}.
Since the extra-dimensional axion is not associated with the cosmological PQ phase transition, as discussed in ref.~\cite{Reece:2023czb}, we can consider it within the canonical pre-inflationary scenario where the PQ symmetry is spontaneously broken during inflation. 
The low-energy effective Lagrangian of the QCD axion, stemming from QCD non-perturbative effects, can be described by
\begin{eqnarray}
\mathcal{L}\supset\dfrac{1}{2} f_a^2 \left(\partial\theta\right)^2-m_a^2 f_a^2\left[1-\cos\left(\theta\right)\right]\, ,
\end{eqnarray}
where $\theta=\phi/f_a$ is the QCD axion angle, $\phi$ represents the QCD axion field, $m_a$ and $f_a$ denote the axion mass and decay constant, respectively. 
Notice that the QCD axion mass is temperature-dependent, which is given by
\begin{eqnarray}
m_a=
\begin{cases}
m_{a,0}\, , & T\leq T_{\rm QCD}\\ 
m_{a,0}\left(\dfrac{T}{T_{\rm QCD}}\right)^{-b}\, , & T> T_{\rm QCD} 
\end{cases} 
\label{mQCDT}
\end{eqnarray} 
where $T_{\rm QCD}\simeq150\, \rm MeV$ is the critical temperature of the QCD phase transition, and $b\simeq4.08$ is an index derived from the dilute instanton gas approximation \cite{Borsanyi:2016ksw}.
The mass at temperatures below $T_{\rm QCD}$ is referred to as the zero-temperature QCD axion mass, which is given by
\begin{eqnarray}
m_{a,0}=\dfrac{m_\pi f_\pi}{f_a}\dfrac{\sqrt{m_u m_d}}{m_u+m_d}\, ,
\end{eqnarray}  
where $m_\pi$ and $f_\pi$ represent the mass and decay constant of the pion, respectively, $m_u$ and $m_d$ are the up and down quark masses.

At high cosmic temperatures, the QCD axion field remains frozen at its initial misalignment angle $\theta_i$.
As the temperature decreases, the axion field begins to oscillate when its mass becomes comparable to the Hubble parameter
\begin{eqnarray}
3H(T_{i,a})=m_a(T_{i,a})\, ,
\end{eqnarray}
where $T_{i,a}$ is the oscillation temperature \cite{Li:2023xkn}
\begin{eqnarray}
T_{i,a}\simeq0.96 \, {\rm GeV}\left(\dfrac{g_*(T_{i,a})}{61.75}\right)^{-0.082}\left(\dfrac{f_a}{10^{12}\, \rm GeV}\right)^{-0.16}\, .
\label{Tosca}
\end{eqnarray}
The axion initial energy density at $T_{i,a}$ is given by
\begin{eqnarray}
\rho_{a,i}=\frac{1}{2}m_{a,i}^2 f_a^2 \left\langle\theta_i^2f(\theta_i)\right\rangle\chi\, ,
\end{eqnarray}
where $m_{a,i}$ is the axion mass at $T_{i,a}$, $f(\theta_i)$ is the anharmonicity factor \cite{Lyth:1991ub}, and $\chi\simeq 1.44$ is a numerical factor \cite{Turner:1985si}.
For temperatures $T_0<T<T_{i,a}$, the QCD axion energy density is adiabatic invariant with the comoving number 
\begin{eqnarray}
N_a \equiv \dfrac{\rho_a}{m_a} \mathfrak{a}^3\, ,
\end{eqnarray}
where $T_0$ is the current cosmic microwave background (CMB) temperature, and $\mathfrak{a}$ is the scale factor.
The axion energy density at $T_0$ is given by
\begin{eqnarray}
\rho_{a,0}=\frac{1}{2}m_{a,0} m_{a,i} f_a^2 \left\langle\theta_i^2f(\theta_i)\right\rangle\chi \left(\frac{\mathfrak{a}_{i,a}}{\mathfrak{a}_0}\right)^3 \, ,
\end{eqnarray}
where $\mathfrak{a}_x$ denotes the scale factor at the temperature $T_x$.
Then the current QCD axion DM abundance can be described by \cite{Li:2023xkn}
\begin{eqnarray}
\Omega_ah^2=\dfrac{\rho_{a,0}}{\rho_{\rm crit}}h^2\simeq0.14 \left(\dfrac{g_{*s}(T_0)}{3.94}\right)\left(\dfrac{g_*(T_{i,a})}{61.75}\right)^{-0.42}\left(\dfrac{f_a}{10^{12}\, \rm GeV}\right)^{1.16}\left\langle\theta_i^2f(\theta_i)\right\rangle\, ,
\end{eqnarray}
where $h\simeq0.68$ is the reduced Hubble constant, $\rho_{\rm crit}$ is the critical energy density, $g_*(T)$ and $g_{*s}(T)$ are the numbers of effective degrees of freedom for the energy density and the entropy density, respectively.
In order to explain the observed DM abundance, $\Omega_{\rm DM}h^2\simeq0.12$ \cite{Planck:2018vyg}, we require an $\sim\mathcal{O}(1)$ initial misalignment angle \cite{Li:2023xkn}
\begin{eqnarray}
\theta_i\simeq0.87\left(\dfrac{g_{*s}(T_0)}{3.94}\right)^{-1/2}\left(\dfrac{g_*(T_{i,a})}{61.75}\right)^{0.21}\left(\dfrac{f_a}{10^{12}\, \rm GeV}\right)^{-0.58}\, .
\end{eqnarray}
Now, considering the QCD axion DM in the dark dimension scenario with a narrow axion decay constant $f_a$ ranging from $10^{9}$ to $10^{10}\, \rm GeV$, we can obtain a small fraction of the overall DM density \cite{Gendler:2024gdo}
\begin{eqnarray}
\dfrac{\Omega_ah^2}{\Omega_{\rm DM}h^2}\sim 10^{-3}-10^{-2}\, ,
\end{eqnarray} 
where the initial misalignment angle $\theta_i$ is assumed to be of order one.  

Then, the situation starts to take a captivating turn.
This is because, in general, the issue we face is the overproduction of the QCD axion DM abundance. 
Specifically, when the decay constant of the QCD axion exceeds $10^{12} \, \rm GeV$ and we only consider an initial misalignment angle of order unity, the axion DM abundance will be overproduced. 
Therefore, in this scenario, fine-tuning the initial misalignment angle emerges as a viable resolution.
However, as analyzed above, when considering the QCD axion in the dark dimension scenario, the value of the axion decay constant within the range of $\sim10^{9}-10^{10}\, {\rm GeV}$ is inadequate to explain the observed DM abundance.
Therefore, our subsequent endeavor will be to explore how to enhance the QCD axion DM abundance for a relatively smaller decay constant $f_a\sim10^{9}-10^{10}\, {\rm GeV}$.
Specifically, we will consider the effect of two-axion mass mixing to address this issue \cite{Cyncynates:2023esj, Li:2024okl}.

\subsection{Enhanced axion abundance through resonant conversion}

In this subsection, we investigate the enhancement of the QCD axion DM abundance within the context of a simple two-axion mixing scenario.
In particular, we consider the mixing between one QCD axion and one ALP.
The low-energy effective Lagrangian that describes this two-axion mixing can be formulated as follows
\begin{eqnarray}
\mathcal{L}\supset\dfrac{1}{2} f_a^2 \left(\partial\theta\right)^2 + \dfrac{1}{2} f_A^2 \left(\partial\Theta\right)^2 -V_{\rm mix}\, ,
\end{eqnarray}
with the general mixing potential
\begin{eqnarray}
V_{\rm mix}=m_a^2 f_a^2\left[1-\cos\left(n_{11}\theta+n_{12}\Theta\right)\right]+m_A^2 f_A^2\left[1-\cos\left(n_{21}\theta+n_{22}\Theta\right)\right]\, ,
\end{eqnarray}
where $\Theta=\varphi/f_A$ is the ALP angle, $\varphi$ represents the ALP field, $m_A$ and $f_A$ denote the ALP mass and decay constant, respectively, and $n_{ij}$ are
the domain wall numbers. 
Notice that the single-field ALP is considered as the simplest case with a constant mass $m_A$.
For our purpose, here the domain wall numbers should be taken as\footnote{The mixing potential for this particular selection was first explored in ref.~\cite{Cyncynates:2023esj}, where it was used to study the mixing between the QCD axion and sterile axion, ultimately resulting in the heavy QCD axion as DM. Additionally, for different selections of domain wall numbers, the axion mixing potential can lead to a suppression of the QCD axion abundance, see $\rm e.g.$ refs.~\cite{Kitajima:2014xla, Daido:2015cba, Ho:2018qur, Li:2023det, Li:2023xkn, Li:2023uvt, Murai:2024nsp}.}
\begin{eqnarray}
n_{11}=n_{12}=n_{22}=1\, , \quad n_{21}=0\, .
\end{eqnarray}
Given that the axion oscillation amplitudes are significantly smaller compared to their respective decay constants, the mass mixing matrix can be described by
\begin{eqnarray}
\mathbf{M}^2=
m_a^2 \left(\begin{array}{cc}
1  & ~ \dfrac{f_a}{f_A}\\
\dfrac{f_a}{f_A} & ~ \left(\dfrac{f_a}{f_A}\right)^2
\end{array}\right)+
\left(\begin{array}{cc}
0  & ~ 0\\
0  & ~ m_A^2
\end{array}\right)\, .
\end{eqnarray}
By diagonalizing the mass mixing matrix, we can derive the heavy ($a_h$) and light ($a_l$) mass eigenstates 
\begin{eqnarray}
\left(\begin{array}{c}
a_h \\
a_l 
\end{array}\right)
=\left(\begin{array}{cc}
\cos \alpha & \quad \sin \alpha \\
-\sin \alpha  & \quad   \cos \alpha
\end{array}\right)
\left(\begin{array}{c}
\phi \\
\varphi 
\end{array}\right)\, ,
\end{eqnarray} 
which represent the two distinct axion states arising from the mixing process.
These eigenstates are associated with the mass eigenvalues $m_{h,l}(T)$, respectively.

\begin{figure}[t]%%%%%%%%%%%%%%%%%%%%%%%%%%TR
\centering
\includegraphics[width=0.70\textwidth]{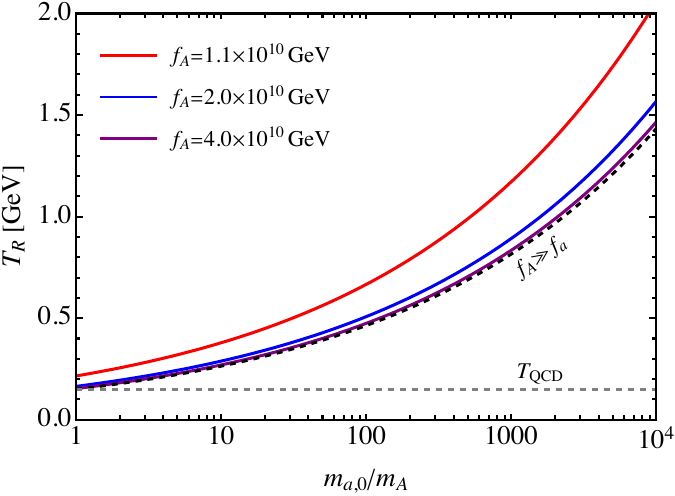}
\caption{The resonant conversion temperature $T_R$ as a function of the axion mass ratio $m_{a,0}/m_A$.
Here we set $f_a=1.0\times10^{10}\, \rm GeV$.
The red, blue, and purple solid lines represent scenarios with $f_A=1.1\times10^{10}\, \rm GeV$, $2.0\times10^{10}\, \rm GeV$, and $4.0\times10^{10}\, \rm GeV$, respectively.
Notice that $f_A$ cannot be equal to $f_a$ here.
The black dashed line corresponds to the case where $f_A$ is much greater than $f_a$.
The gray dashed line represents the QCD phase transition critical temperature $T_{\rm QCD}$.}
\label{fig_TR}
\end{figure}

Next we discuss the axion resonant conversion within the context of this specific mixing.
For our analysis, we consider the general prerequisites for resonant conversion, which are expressed as follows
\begin{eqnarray}
\dfrac{m_{a,0}}{m_A}\gg 1 \, , \quad  \dfrac{f_a}{f_A}\ll 1\, .
\end{eqnarray} 
The resonant conversion temperature $T_R$ is given by solving $d\left(m_h^2(T)-m_l^2(T)\right)/dT=0$, yielding the precise expression
\begin{eqnarray}
T_R=T_{\rm QCD}\left(\dfrac{m_{a,0}^2\left(f_a^2+f_A^2\right)^2}{m_A^2 f_A^2\left(f_A^2-f_a^2\right)}\right)^{\frac{1}{2b}}\, .
\end{eqnarray} 
In the scenario where $f_A\gg f_a$, it is possible to approximate the temperature $T_R$ such that the value of $m_a(T_R)$ is approximately equal to $m_A$, leading to
\begin{eqnarray}
T_R\simeq T_{\rm QCD}\sqrt[b]{\dfrac{m_{a,0}}{m_A}}\, .
\end{eqnarray}  
See figure~\ref{fig_TR} for the distribution of $T_R$ as a function of the axion mass ratio $m_{a,0}/m_A$, with the QCD axion decay constant set at $f_a=1.0\times10^{10}\, \rm GeV$.
This figure highlights the substantial variations in $T_R$ as the ALP decay constant $f_A$ increases.
Notably, we can observe that the resonant conversion temperature demonstrates an asymptotic behavior when $f_A$ is marginally greater than $f_a$, indicating that a significant disparity between $f_A$ and $f_a$ is not prerequisite for this distribution to emerge.

At temperatures above $T_R$, the heavy mass eigenstate $a_h$ primarily consists of the ALP, while the light mass eigenstate $a_l$ is dominated by the QCD axion.
As the cosmic temperature decreases, the mass eigenvalues $m_h(T)$ and $m_l(T)$ converge at $T = T_R$ and then diverge.
Below the temperature $T_R$, $a_h$ becomes the QCD axion, and $a_l$ transforms into the ALP.
Here, the axion energy transition at $T_R$ is presumed to be adiabatic, which is approximately valid when \cite{Cyncynates:2023esj}
\begin{eqnarray}
T_R \ll T_{i,a}\, ,
\label{adiabatic_app}
\end{eqnarray}
where $T_{i,a}$ represents the QCD axion oscillation temperature shown in eq.~\eqref{Tosca}.
In this case, the ALP begins oscillating prior to $T_{i,a}$.
Subsequently, it is crucial to examine the range of the ALP mass $m_A$, which can be characterized by
\begin{eqnarray}
m_{a,i} < m_A < m_{a,0}\, ,
\label{ALP_mass_relation}
\end{eqnarray}
where $m_{a,i}\equiv m_a(T_{i,a})$.
Therefore, for distinct values of the QCD axion decay constant $f_a = 10^9$ and $10^{10} \, \rm GeV$, respectively, we obtain the following ALP mass ranges
\begin{eqnarray}
m_A\sim 10^{-6}-10^{-3}\, {\rm eV}\, , ~ m_A\sim 10^{-7}-10^{-4}\, {\rm eV}\, .
\label{ALP_mass_range}
\end{eqnarray}
To satisfy both sub-formulas simultaneously, we will choose a typical value of $m_A=10^{-5}\, \rm eV$ for the ALP mass in our subsequent discussion. 
Since the ALP mass is relatively flexible, this choice is reasonable.

\begin{figure}[t]%%%%%%%%%%%%%%%%%%%%%%%%%%mAfA
\centering
\includegraphics[width=0.92\textwidth]{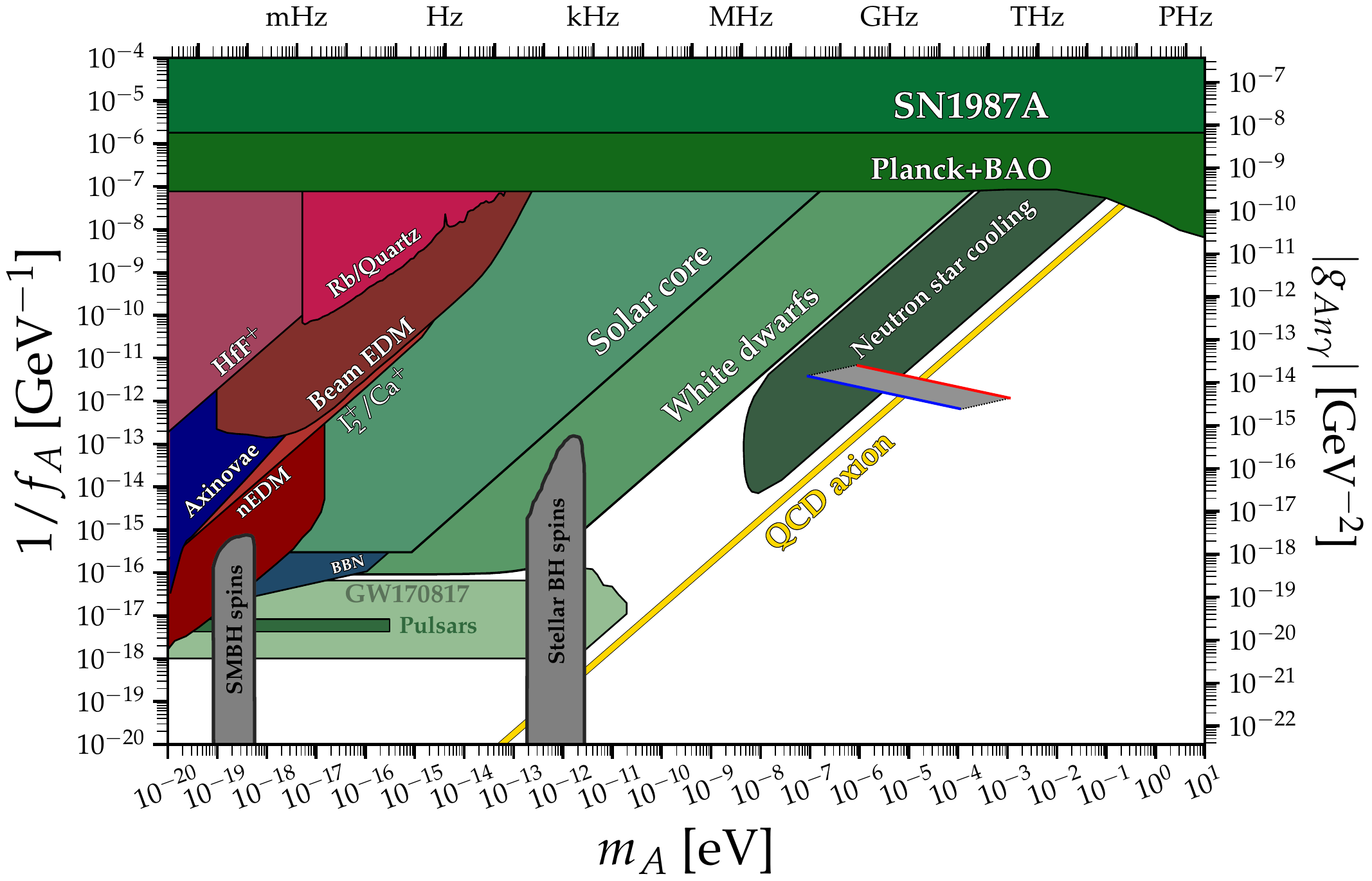}
\caption{The ALP constraints in the $\{m_A, 1/f_A\}$ plane.
The red and blue solid lines correspond to the scenarios
where the QCD axion decay constant is $f_a=10^{9}\, \rm GeV$ and $10^{10}\, \rm GeV$, respectively.
The shaded area between these two lines represents the allowed ALP parameter space that can account for the overall DM abundance in our scenario.
Additional constraints on ALP within this plane are sourced from ref.~\cite{ciaran_o_hare_2020_3932430}.}
\label{fig_mAfA}
\end{figure}

In order to determine the abundance of QCD axion DM in the context of axion mixing, we should begin with the ALP field. 
At high temperatures, the ALP field is frozen at its initial misalignment angle, $\Theta_i$. 
It begins to oscillate at the temperature $T_{i,A}$ with the initial energy density
\begin{eqnarray}
\rho_{A,i}=\frac{1}{2}m_A^2 f_A^2 \left\langle\Theta_i^2f(\Theta_i)\right\rangle \, ,
\end{eqnarray}
where $T_{i,A}$ is given by $3H(T_{i,A})=m_A$.
For temperatures $T_R < T < T_{i,A}$, the ALP energy density is adiabatic invariant with the comoving number $N_A \equiv \rho_A \mathfrak{a}^3 /m_A$. 
At $T_R$, this energy density is expressed as
\begin{eqnarray}
\rho_{A,R}=\frac{1}{2}m_A^2 f_A^2 \left\langle\Theta_i^2f(\Theta_i)\right\rangle \left(\frac{\mathfrak{a}_{i,A}}{\mathfrak{a}_R}\right)^3 \, ,
\end{eqnarray}
where $\mathfrak{a}_{i,A}$ and $\mathfrak{a}_R$ are the scale factors at $T_{i,A}$ and $T_R$, respectively. 
Then the energy density of the ALP, $\rho_{A,R}$, is transferred to the QCD axion, $\rho_{a,R}$.
For temperatures below $T_R$, the QCD axion energy density is adiabatic invariant.
At $T_0$, this energy density is given by
\begin{eqnarray}
\rho'_{a,0}=\frac{1}{2}m_{a,0}m_A f_A^2 \left\langle\Theta_i^2f(\Theta_i)\right\rangle \left(\frac{\mathfrak{a}_{i,A}}{\mathfrak{a}_0}\right)^3 \, ,
\end{eqnarray}
where $T_0$ is the present CMB temperature, and $\mathfrak{a}_0$ is the scale factor at $T_0$.
Comparing this with the case without mixing, we find that the QCD axion energy density can be modified by a factor
\begin{eqnarray}
R_\rho \equiv \dfrac{\rho'_{a,0}}{\rho_{a,0}}\, .
\end{eqnarray}
Now, given the initial misalignment angles ($\theta_i$ and $\Theta_i$) of $\sim\mathcal{O}(1)$, and in order to account for the total DM density with $f_a\sim 10^{9}-10^{10}\, \rm GeV$, we simply require that
\begin{eqnarray}
R_\rho \sim 10^{2}-10^{3}\, .
\end{eqnarray}
In this case, the QCD axion DM abundance can be approximated as follows
\begin{eqnarray}
\begin{aligned}
\Omega'_a h^2&\simeq0.17 \left(\dfrac{g_{*s}(T_0)}{3.94}\right)\left(\dfrac{g_*(T_{i,A})}{61.75}\right)^{-1/4}\\
&\times \left(\dfrac{f_a}{10^{10}\, \rm GeV}\right)^{-1} \left(\dfrac{f_A}{10^{12}\, \rm GeV}\right)^{2} \left(\dfrac{m_A}{10^{-5}\, \rm eV}\right)^{-1/2} \left\langle\Theta_i^2f(\Theta_i)\right\rangle\, .
\label{abundance_mixing}
\end{aligned}
\end{eqnarray}
By assuming a typical ALP mass of $m_A=10^{-5}\, \rm eV$, and considering the observed cold DM abundance $\Omega_{\rm DM}h^2\simeq0.12$ with the QCD axion in the dark dimension, where $f_a\sim 10^{9}-10^{10}\, \rm GeV$, we find that the ALP decay constant falls within the range    
\begin{eqnarray}
f_A\sim 2.7\times 10^{11}-8.4\times 10^{11}\, \rm GeV\, .
\label{ALP_decay_constant}
\end{eqnarray}
Utilizing eqs.~\eqref{ALP_mass_range} and \eqref{abundance_mixing}, we also show in figure~\ref{fig_mAfA} the allowed ALP parameter space in the $\{m_A, 1/f_A\}$ plane.
The red and blue solid lines represent the scenarios where $f_a=10^{9}\, \rm GeV$ and $10^{10}\, \rm GeV$, respectively.
The shaded gray quadrilateral area bounded by these two lines represents the allowed ALP parameter space within our scenario, encompassing the range that accounts for the overall DM abundance via axion resonant conversion.

Several aspects require further clarification and elaboration for a clear understanding.
Regarding eq.~\eqref{ALP_mass_range}, there are two key points to clarify.
Firstly, it is presented considering that in certain regions, the ALP will no longer be lighter than the QCD axion.
Secondly, we have considered keeping the resonant conversion temperature as low as possible compared to the axion oscillation temperature, to ensure that adiabatic conversion can occur completely.\footnote{We have assumed the approximation that all ALP density is transferred to the QCD axion, which is acceptable for large disparities in axion decay constants. When the adiabatic condition is violated, this resonant conversion does not occur completely, and thus the allowed parameter space for ALP may expand. Regarding the extent to which adiabatic condition is violated, this remains uncertain and requires further complex numerical analysis. At least, this condition should not be violated in this work, since $f_a/f_A\sim 10^{-3}-10^{-2}\ll 1$. Nevertheless, the constraints presented here conservatively determine the ALP parameter space in the case of complete conversion.}
These two points are addressed in eqs.~\eqref{ALP_mass_relation} and \eqref{adiabatic_app}, which provide the mass range of ALP in eq.~\eqref{ALP_mass_range}.
Regarding eq.~\eqref{abundance_mixing}, it involves three variables: $f_a$, $f_A$, and $m_A$. 
To illustrate the ALP parameter space in the $\{m_A, 1/f_A\}$ plane, we need to fix the value of $f_a$. 
Consequently, there will be two lines corresponding to $f_a=10^{9}\, \rm GeV$ and $10^{10}\, \rm GeV$, represented as the red and blue lines in the figure, respectively. 
Naturally, the range of these red and blue lines is also constrained by eq.~\eqref{ALP_mass_range}. 
Then the shaded area between these two lines represents the allowed region for the ALP that we aim to obtain in this work.
Furthermore, we also find that a significant portion of ALPs falling within this specific range have the potential to be detected by future axion experiments.

Additionally, in the context of considering a scenario with multiple QCD axions mass mixing \cite{Li:2024okl}, the abundance of QCD axion DM can also be effectively enhanced, depending on the relationship among axion decay constants. 
This scenario may also provide an explanation for the issue of QCD axion DM abundance in the dark dimension scenario, requiring further investigation.

\subsubsection{Brief discussion on adiabatic condition} 

In this subsection, we briefly discuss the adiabatic condition in axion resonant conversion.
The previous discussion roughly considered the adiabatic condition to be satisfied when the temperature at which axions begin to oscillate is much higher than the resonant conversion temperature, $\rm i.e.$, $T_{i,a} \gg T_R$.
Actually, the precise form of the adiabatic condition was first given in ref.~\cite{Ho:2018qur}, and its expression is as follows, $\Delta t_R \gg \max [2\pi/ m_l(T_R), \, 2\pi/(m_h(T_R)-m_l(T_R))]$, where $\Delta t_R$ represents the duration of the resonant conversion period.
It implies that the comoving axion numbers of the eigenstates $a_h$ and $a_l$ are individually conserved at the temperature $T_R$.

Very recently, ref.~\cite{Murai:2024nsp} further discussed the adiabatic condition within the context of resonant conversion between the QCD axion and ALP.
They clarified the relationship among different bases used to describe the axion resonant conversion and derived an improved, basis-independent definition of the adiabatic condition, thereby addressing the limitations inherent in previous formulations.

\subsection{Possible origins of ALP with a higher-scale decay constant} 

In this subsection, we discuss the possible origins of the ALP with a higher-scale decay constant that is required to satisfy the preceding eq.~\eqref{ALP_decay_constant}. 
We speculate that the ALP, which falls within this specific range, may originate from the grand unification of gauge forces in the context of the dark dimension, where realizing such unification is highly constrained \cite{Heckman:2024trz}.
In this scenario, the WGC constraint on the ALP decay constant is modified by a factor of $M_5^{1/2} L_g^{1/2}\sim 10^3$, resulting in
\begin{eqnarray}
f_A\lesssim 10^3\times M_5\sim 10^{12}-10^{13}\, \rm GeV\, ,
\end{eqnarray}
which sits comfortably in the permitted experimental range, where $L_g$ represents the diameter of the brane accommodating gauge fields,
\begin{eqnarray}
L_g\lesssim \left(1-10\, {\rm TeV}\right)^{-1}\, .
\end{eqnarray} 
However, in this case, the QCD axion may also be subject to this constraint, leading to a larger decay constant too. 
This, in turn, may reduce the need for the ALP resonant conversion discussed here, although having a more weakly-coupled axion might be less appealing.
Unless, of course, the decay constant of the QCD axion in this scenario is also subject to other unknown constraints. 

Specifically, it is natural for us to further consider the framework of what is termed the ``dark-dimensional axiverse,$"$ where both the QCD axion and a large number of ALPs coexist, leading to the Lagrangian
\begin{eqnarray}
\mathcal{L}\supset\dfrac{1}{2}\sum_i f_i^2 \left(\partial a_i \right)^2-\sum_i\Lambda_i^4\left[1-\cos\left(\sum_j n_{ij} a_j \right)\right]\, ,
\end{eqnarray} 
where $a_i$, $f_i$, $\Lambda_i$, and $n_{ij}$ represent the axion angles, decay constants, overall scales, and domain wall numbers, respectively.
Within such a model, the decay constant of the QCD axion can be set at a relatively small value, whereas the decay constant of some ALPs may be significantly enhanced. 
Notice also that this type of multi-axion scenario, with hierarchical axion decay constants, can be realized in string theory, such as the commonly encountered type IIB string axiverse \cite{Cicoli:2012sz}.
Nevertheless, this exceeds the scope of the present paper and merits further exploration in future work.

\section{Conclusion}%%%%%%%%%%%%%%%%%%%%%%%%%%%Conclusion 
\label{sec_Conclusion}

In summary, we have investigated the issue of QCD axion DM within the context of the recently proposed dark dimension scenario, and demonstrated that the QCD axion can account for the overall DM abundance through the axion mass mixing mechanism.
We first provide a concise overview of the QCD axion in the dark dimension. 
Subsequently, we discuss the QCD axion DM in this scenario through the misalignment mechanism.
Then we discuss the enhancement of the axion DM abundance within the context of a simple two-axion mixing scenario.

The dark dimension scenario is motivated by the minuscule nature of dark energy, informed by Swampland principles and constrained by observational data.
In this context, axions can be localized on the SM brane.
By applying the WGC to the QCD axion, one can derive an upper bound for the axion decay constant, $f_a\lesssim M_5 \sim 10^{9}-10^{10}\, \rm GeV$, where $M_5$ is the five-dimensional Planck mass.  
When combined with observational lower bounds, this suggests that $f_a$ falls within the range $f_a\sim 10^{9}-10^{10}\, \rm GeV$, corresponding to an axion mass $m_a\sim 10^{-3}-10^{-2}\, \rm eV$. 
According to the misalignment mechanism, the QCD axion at this scale could constitute a minor fraction of the total DM density $\sim 10^{-3}-10^{-2}$ if without the fine-tuning of the initial misalignment angle. 
In order to enhance the QCD axion DM abundance, we consider a straightforward two-axion mixing mechanism, specifically the resonant conversion of an ALP into the QCD axion.
Our findings suggest that, in a scenario where the ALP possesses a mass of approximately $m_A \sim 10^{-5} \, \rm eV$ and a decay constant of $f_A \sim 10^{11} \, \rm GeV$, the QCD axion in the dark dimension scenario can account for the overall DM abundance.
This is supported by a detailed analysis of the axion mixing potential and the temperature effects on the axion energy density.
We also speculate that the ALP required within this specific range of decay constant may stem from the grand unification in the context of the dark dimension.
Furthermore, future experiments aimed at detecting ALPs within this range have the potential to further validate this scenario. 
Additionally, a scenario with multiple QCD axions mass mixing should also be able to explain the issue discussed in this work.
Our work highlights the potential of the QCD axion within the dark dimension scenario to account for the cold DM abundance.
 
\section*{Acknowledgments}%%%%%%%%%%%%%%%%Acknowledgments

We thank Michele Cicoli and Yu-Feng Zhou for useful discussions.
This work was supported by the Key Laboratory of Theoretical Physics in Institute of Theoretical Physics, CAS.

%\appendix

%\section{}

\bibliographystyle{JHEP}
\bibliography{references}

\end{document}